# Generalized Retrieval Method for Metamaterial Constitutive Parameters Based on a Physically-Driven Homogenization Approach

Xing-Xiang Liu[†], and Andrea Alù[*]

Department of Electrical and Computer Engineering, The University of Texas at Austin,

Austin, TX 78712, U.S.A.

**Abstract**

Based on the recently introduced homogenization theory developed in [Phys. Rev. B 84, 075153 (2011)], we propose a generalized retrieval method that allows extracting physically meaningful bulk effective parameters from conventional scattering measurements of periodic metamaterial samples composed of subwavelength inclusions. We show that, compared to conventional approaches, our method is able to capture the anomalous physics in the wave interaction with resonant metamaterials and return physically meaningful homogenized parameters that retain local properties in the long-wavelength limit. As a byproduct, we are also able to retrieve the polarizabilities of the constituent inclusions, which are shown to satisfy expected dispersion properties for passive inclusions, in contrast with conventional retrieval approaches.

[†] E-mail: xxliu@utexas.edu

[*]To whom correspondence should be addressed. E-mail: alu@mail.utexas.edu

## I. Introduction

Determining the effective constitutive parameters of a metamaterial sample is essential to describe its bulk wave interaction in a macroscopic sense. Being able to define physically meaningful permittivity and permeability of such composite materials made of collections of resonant, often complex inclusions, such as split-ring resonantors[1] and omega-shaped elements[2], would allow to simplify its electromagnetic description by neglecting the deeply subwavelength field details of its microstructure and to conceptually treat it as a homogenous material.

Given a metamaterial sample, it is common to apply well established retrieval methods for conventional materials[3-6], such as the Nicolson-Ross-Weir (NRW) technique, to determine its homogenized constitutive parameters. The idea is essentially to assume a homogeneous bulk model and assign constitutive parameters that fit the observed scattering properties. This retrieval approach requires exciting the sample with specific wave forms (usually plane waves) and solving an inverse problem to find the equivalent bulk parameters that would provide the same scattering responses. This procedure is very useful for conventional materials, but it often presents challenges in the case of metamaterials. Typically retrieved parameters have been shown to violate conventional causality and passivity conditions for permittivity and permeability[7-9], due to over-simplified initial assumptions on the bulk model. This issue becomes especially important near the inclusion resonances that support the interesting electromagnetic response of metamaterials. In addition, when the metamaterial samples under analysis are electrically thick, determining the correct branches for the propagation constant inside the array becomes ambiguous[4]. Environmental noise in scattering measurements and numerical error in computations, even at extremely low levels, may be largely amplified near resonances and create additional nonphysical artifacts[10].

These limitations of conventional retrieval methods have stimulated the development of refined homogenization procedures especially tailored for metamaterials[11-20] that aim at capturing with more precision and accuracy the complex wave interaction in these artificial media. These studies have been able to explain some of the reasons behind the nonphysical features of commonly retrieved metamaterial parameters, and have highlighted effects, such as spatial dispersion and bianisotropic response, that are usually neglected in simple retrieval methods and lie at the roots of their limitations[21]. Recently, we have put forward a first-principle homogenization theory (FPHT)[20,21] that properly takes into account weak forms of spatial dispersion in periodic metamaterials and determines physically meaningful bulk parameters in the long-wavelength limit. This approach is a direct homogenization method, in which the homogenized effective parameters are obtained based on the known microscopic details of the metamaterial under analysis. No inverse retrieval procedure was proposed however to solve the arguably more interesting problem of extracting the proper parameters from a given metamaterial sample. Following this theory, the authors of Ref. 22 have proposed a suitable correction to conventional NRW retrieval that may provide more meaningful constitutive parameters with a causal Lorentzian dispersion in frequency. However, a complete and systematic retrieval method to derive physically meaningful parameters as rigorously defined in the FPHT approach has not been put forward yet.

This improved procedure is the scope of the present work, in which we develop a generalized retrieval method (GRM) based on the FPHT to evaluate physically meaningful effective parameters given the scattering properties of a metamaterial sample. The proposed method is divided in two steps: we first employ an improved version of the conventional NRW retrieval to determine *equivalent* constitutive parameters, as defined in Refs. 10 and 20, with the same limitations as described above. Secondly, we apply the FPHT results to restore the physical meaning of these parameters and derive the *effective* constitutive parameters of the metamaterial, properly taking into account weak forms of spatial dispersion neglected in conventional NRW retrieval approaches[23]. The proposed GRM not only determines physically meaningful bulk parameters for a broad class of metamaterial geometries, but also returns, as a byproduct, the polarizabilities of each unit cell, which capture the scattering characteristics of the elementary inclusions composing the array.

After formulating the GRM, we apply it to a representative example, a metamaterial composed of subwavelength magnetodielectric spheres, and we compare the conventional NRW approach with this method to discuss physical insights and highlight its advantages. We demonstrate that the GRM returns effective constitutive parameters that satisfy the expected causality relations, in contrast with conventional retrieval approaches, and we also show its inherent robustness to measurement errors, associated with the physical foundations of its formulation. In addition to the physical insights presented in the following, from the engineering point of view this technique becomes a valuable tool to derive meaningful homogenized parameters with local properties, which can therefore be applied to different metamaterial shapes and forms of excitation.

## II.   Theoretical formulation

In this work, we formulate the GRM for a general periodic metamaterial array composed of subwavelength inclusions in an orthorhombic lattice. Extension to different lattice configurations would be straightforward, but it is not an objective of this paper. Provided that the subwavelength inclusions are not too closely packed, their scattering properties and collective interaction can be described by the dipolar polarizabilities $\alpha_e$ (electric) and $\alpha_m$ (magnetic), which are assumed scalar here to keep the formulation simpler. For the sake of simplicity, and to better highlight the merits of the proposed retrieval method compared to conventional techniques, we also assume absence of magneto-electric coupling within the inclusions, i.e., $\alpha_{em} = \alpha_{me} = 0$. These assumptions effectively imply specific forms of symmetry in the inclusions within the lattice with respect to the impinging polarization, but an extension to a fully generalized tensorial formulation including chiral effects is possible following a similar scheme, and will be presented in the near future.

Following Ref. 20, we may describe this periodic metamaterial with physically meaningful constitutive relations in the general form

$$\begin{aligned} \mathbf{D}_{av} &= \varepsilon_0 \mathbf{E}_{av} + \mathbf{P}_{av} = \varepsilon_{eff} \mathbf{E}_{av} - \chi_{eff}^o \hat{\boldsymbol{\beta}} \times \mathbf{H}_{av} \\ \mathbf{B}_{av} &= \mu_0 \mathbf{H}_{av} + \mathbf{M}_{av} = \mu_{eff} \mathbf{H}_{av} + \chi_{eff}^o \hat{\boldsymbol{\beta}} \times \mathbf{E}_{av} \end{aligned}, \quad (1)$$

where the subscript *av* denotes properly averaged fields within a unit cell, as defined in Ref. 24, $\hat{\boldsymbol{\beta}}$ is the propagation unit vector in the metamaterial, and the *effective* constitutive parameters may be written in a generalized Clausius-Mossotti closed form as

$$\varepsilon_{eff} = \varepsilon_0 \left[ 1 + \frac{1}{d^3} \frac{\alpha_m^{-1} - C_{int}}{\left( \alpha_e^{-1} - C_{int} \right)\left( \alpha_m^{-1} - C_{int} \right) - C_{em}'^2} \right], \quad (2)$$

$$\mu_{eff} = \mu_0 \left[ 1 + \frac{1}{d^3} \frac{\alpha_e^{-1} - C_{int}}{\left( \alpha_e^{-1} - C_{int} \right)\left( \alpha_m^{-1} - C_{int} \right) - C_{em}'^2} \right], \quad (3)$$

$$\chi_{eff}^o = \frac{1}{c_0 d^3} \frac{C_{em}'}{\left( \alpha_e^{-1} - C_{int} \right)\left( \alpha_m^{-1} - C_{int} \right) - C_{em}'^2}. \quad (4)$$

Here we have used the interaction coefficients

$$C_{\text{int}} = -\frac{1}{d^3}\frac{k_0^2}{\beta^2 - k_0^2} + \sum_{(l,m,n)\neq(0,0,0)} \hat{\mathbf{p}}\cdot\mathbf{G}_{ee}(\mathbf{r}_{lmn})e^{i\boldsymbol{\beta}\cdot\mathbf{r}_{lmn}}\cdot\hat{\mathbf{p}}, \tag{5}$$

$$C'_{em} = -\left[\frac{1}{d^3}\frac{\beta k_0}{\beta^2 - k_0^2}\right] + \sum_{(l,m,n)\neq(0,0,0)} \hat{\mathbf{p}}\cdot\mathbf{G}_{em}(\mathbf{r}_{lmn})e^{i\boldsymbol{\beta}\cdot\mathbf{r}_{lmn}}\cdot\hat{\mathbf{m}}, \tag{6}$$

describing the co- and cross-polarization coupling of electric and magnetic dipole moments in each unit cells based on Green's functions $\mathbf{G}_{ee}$ and $\mathbf{G}_{em}$. In order to efficiently calculate the infinite summations involved in these interaction coefficients, fast algorithms are required[11,25]. In addition, $\beta = |\boldsymbol{\beta}|$ and $k_0 = \omega/c_0$ are respectively the guided and free-space wave number in an arbitrary $e^{i(\boldsymbol{\beta}\cdot\mathbf{r}-\omega t)}$ Floquet-based space-time dependence of the averaged fields, and $\varepsilon_0, \mu_0, c_0$ are the free-space permittivity, permeability and speed of light, respectively.

It is important to note that the constitutive relations (1) are written in the general Tellegen (or bianisotropic) form[26], which implies a form of magneto-electric coupling relating $\mathbf{D}_{av}$ to $\mathbf{H}_{av}$ and $\mathbf{B}_{av}$ to $\mathbf{E}_{av}$, even in the present case for which the inclusions in the lattice are center-symmetric. This is due to the presence of the bianisotropy parameter $\chi^o_{eff}$. As discussed in our previous works[21,27], this inherent form of bianisotropy, especially relevant near the inclusion or lattice resonances, is produced by weak nonlocal effects arising even in the long wavelength limit when the granularity of the array is not negligible compared to $1/\beta$. By properly taking into account these effects, it is possible to restore local and physically meaningful effective parameters[21] in the long-wavelength regime.

It should be stressed that the homogenized description in Eq. (1) is valid for arbitrary excitation of the structure and not only in its eigen-modal operation, as proven in Ref. 21. This property is very appealing for practical applications, as it allows using these parameters independent on the form and type of excitation of the metamaterial sample. In the general case, the constitutive parameters (2)-(4) are functions of the independent variables $\beta$ and $k_0$. However, in the special case in which no sources are impressed inside the metamaterial, the values of $\beta$ and $k_0$ are inherently related to each other through the eigen-modal dispersion relation, which also forces a relation between the averaged electric and magnetic fields and allows rewriting Eq. (1) in the more common form

$$\begin{aligned}\mathbf{D}_{av} &= \varepsilon_{eq}\mathbf{E}_{av}\\ \mathbf{B}_{av} &= \mu_{eq}\mathbf{H}_{av}\end{aligned}, \quad (7)$$

with these vectors satisfying the usual macroscopic form of Maxwell's equations. There is a direct relationship between the *effective* parameters used in (1) and the *equivalent* parameters in (7), which may be easily derived from Maxwell's equations[20]:

$$\varepsilon_{eq} = \frac{\varepsilon_{eff}}{1 - \dfrac{c_0 \chi_{eff}^o}{\beta/k_0}}, \quad (8)$$

$$\mu_{eq} = \frac{\mu_{eff}}{1 - \dfrac{c_0 \chi_{eff}^o}{\beta/k_0}}. \quad (9)$$

Both constitutive models are valid in the absence of embedded sources if the goal is simply to describe the array propagation and bulk scattering properties of the metamaterial, and this explains the success of the conventional NRW approach, which is essentially based on the *equivalent* description, to describe and tailor the scattering of metamaterials. However, we need to keep in mind that this *equivalent* description is inherently nonlocal even in the long-wavelength limit, and therefore should not be expected to follow the usual requirements and properties on local permittivity and permeability, nor even share their commonly accepted meaning. In our previous works[21,27], we indeed proved that this is the reason behind the common violation of Kramers-Kronig and causality relations associated with the *equivalent* constitutive parameters, which implicitly hide the weak spatial dispersion effects captured in $\chi_{eff}^{o}$ and are valid only for eigen-modal propagation.

As mentioned, since the constitutive model (7) is the same as the one conventionally assumed in NRW retrieval procedures, the equivalent parameters in Eqs. (8)-(9) are consistent with the parameters $\varepsilon_{NRW}$ and $\mu_{NRW}$ that we would extract by using this standard retrieval method[21,27], and they obviously share the same limitations. Small differences between the analytical formulas in (8)-(9) and the parameters retrieved from a scattering measurement may be expected, due to the dipolar assumptions used in (2)-(4). These differences will be more significant for dense arrays, for which higher-order multipolar scattering plays a more important role.

In the following, we are interested in improving the retrieval procedure to take into account these concepts and propose a way to extract the more physically meaningful effective parameters in (2)-(4). We concentrate on the eigen-modal scenario in which sources are not embedded in the metamaterial sample, which is the more common case and allows a direct comparison with conventional NRW retrieval. In this case, the metamaterial secondary parameters, i.e., the characteristic impedance $\eta$ and the wave number $\beta$, fully characterize the metamaterial bulk response in the limit in which only one eigenmode is supported at the given frequency of excitation. These parameters are related to the three sets of constitutive parameters defined above through the following relations:

$$\eta = \sqrt{\frac{\mu_{NRW}}{\varepsilon_{NRW}}} \approx \sqrt{\frac{\mu_{eq}}{\varepsilon_{eq}}} = \sqrt{\frac{\mu_{eff}}{\varepsilon_{eff}}}, \qquad (10)$$

$$\beta = \omega\sqrt{\mu_{NRW}\varepsilon_{NRW}} \approx \omega\sqrt{\mu_{eq}\varepsilon_{eq}}$$
$$= \omega\sqrt{\varepsilon_{eff}\mu_{eff}} \bigg/ \sqrt{1 - \frac{c_0 \chi_{eff}^o}{\beta/k_0}}, \qquad (11)$$

where the nearly-equal sign indicates the potential difference between NRW and equivalent parameters due to the influence of higher-order multipoles. A conventional retrieval procedure would first derive $\eta$ and $\beta$ from the scattering measurements (usually the reflection $R$ and transmission $T$ from a planar slab) and then obtain the NRW parameters by inverting Eqs. (10)-(11). It is not possible, however, to determine the *effective* parameters using just Eqs. (10)-(11), since $\eta$ and $\beta$ do not univocally determine them. This is related to the fact, discussed in more details in Ref. 20, that $\chi_{eff}^o$ is essentially a measure of the spatial dispersion stemming from the fact that $\beta d$ is not infinitesimally small, and therefore the wave feels the lattice granularity as it propagates across each unit cell. For this reason, different combinations of *effective* parameters can produce the same values of $\eta$ and $\beta$, as a function of $\beta d$. In order to accurately capture these weak spatial dispersion effects, it is not sufficient to just measure $R$ and $T$, but we also need to know the average array period $d$ of the array under analysis. With the knowledge of this geometry parameter, which is easily obtainable for a given metamaterial, we may invert the system formed by the combination of Eqs. (2)-(4) and (10)-(11) to retrieve the effective constitutive parameters of the metamaterial under analysis.

In practice, we have a set of five independent relations that may be solved for the five complex unknowns $\varepsilon_{eff}$, $\mu_{eff}$, $\chi_{eff}^o$, $\alpha_e$ and $\alpha_m$. This implies that the proposed GRM does not only provide the effective parameters, but it also returns the averaged polarizability of each unit cell as a byproduct. Although it is not possible to analytically invert the system due to the complex dependence of the coupling parameters $C_{int}$ and $C'_{em}$ on $\beta$, we may use an optimized numerical procedure or root-finding method to solve for the unknowns. In the following section, we apply this method to a specific metamaterial sample and outline its advantages compared to the conventional retrieval approach.

### III. Effective parameters extracted from the generalized retrieval method

We apply the proposed GRM to extract the effective local parameters of a cubic array with period $d$ of magnetodielectric spheres of radius $a$. The spheres have permittivity $\varepsilon = 13.8\varepsilon_0$ and permeability $\mu = 11.0\mu_0$ and we assume a dense packing factor $a/d = 0.45$ in a free-space background. These parameters are consistent with one of the geometries analyzed in Refs 16,20,28 to realize quasi-isotropic double-negative metamaterials using center-symmetric inclusions that do not have any form of inherent magneto-electric coupling or chirality and are independent of the polarization or direction of incidence.

As discussed in the previous section, we can first apply a conventional NRW retrieval procedure to determine the secondary parameters of the array and the associated *equivalent* constitutive model (8)-(9). We use CST Microwave Studio to numerically compute the reflection and transmission coefficients of a finite-thickness slab composed of 6 layers of unit cells in the propagation direction. The impedance and wave numbers may be easily retrieved from $R$ and $T$ using the following equations, as a function of the transverse-electric (TE) or transverse-magnetic (TM) polarization and incidence angle $\theta_i$ [29]:

$$\text{TE}: \begin{cases} \eta^2 = \eta_0^2 \dfrac{(1+R)^2 - T^2}{(1-R)^2 - T^2}\left(\dfrac{\cos\theta_t}{\cos\theta_i}\right)^2 \\ \cos(\beta d \cos\theta_t) = \dfrac{1 - R^2 + T^2}{2T} \end{cases}, \qquad (12)$$

$$\text{TM}: \begin{cases} \eta^2 = \eta_0^2 \dfrac{(1-R)^2 - T^2}{(1+R)^2 - T^2}\left(\dfrac{\cos\theta_t}{\cos\theta_i}\right)^2 \\ \cos(\beta d \cos\theta_t) = \dfrac{1 - R^2 + T^2}{2T} \end{cases}. \qquad (13)$$

In these formulas, the refraction angle $\theta_t$ depends on the incident angle as $\sin\theta_t = k_0 \sin\theta_i / \beta$.

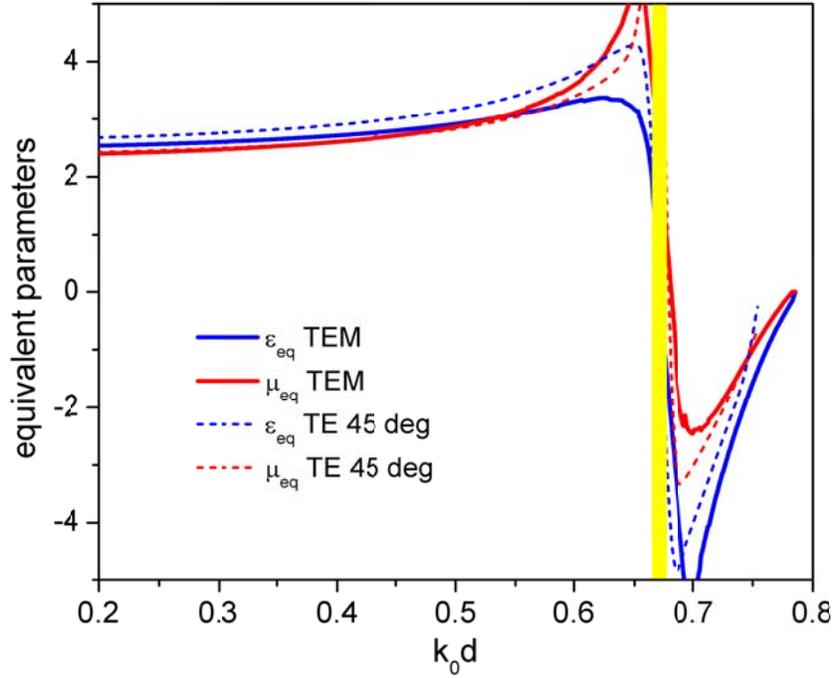

Figure 1 - Equivalent parameters obtained from an improved NRW retrieval consistent with Ref. 10 for normal incidence (solid lines) and 45-degree TE incidence (dashed lines).

Figure 1 shows the conventionally retrieved parameters for normal incidence and for 45-degree TE incidence[30]. In these calculations, we have used an improved procedure, as described in Ref. 10, to suppress numerical artifacts inherently arising near the Fabry-Perot resonances of the slab. It is evident that the retrieved parameters show quite remarkable differences for different incidence angles. These discrepancies become especially important near the bandgap region marked by a yellow shadow in the figure, and are an indication of spatial dispersion in the array. Right past this bandgap region, the metamaterial supports a double-negative operation, in which both equivalent permittivity and permeability are negative, and in this region the relevant differences between the retrieved parameters are also evident for different incidence angles.

The retrieved parameters, as expected, are purely real within the propagation region due to the absence of losses, but they have a negative slope versus frequency around the bandgap region that violates the dispersion expected from Kramers-Kronig relations[31]. This is consistent with the common violation of Kramers-Kronig relations of the retrieved metamaterial parameters, and it is also another indication of the spatial dispersion and nonlocality effects outlined in the previous section. When the metamaterial operates near the inclusion resonance, especially in the negative-index regime, the magneto-electric coupling associated with the finite granularity of the array and taken into account by $\chi_{eff}^{o}$ cannot be captured by the simple constitutive model assumed in this conventional NRW approach. The phase variation across each unit cell cannot be ignored, as implicitly assumed in a simple retrieval procedure, and this is the root of the reason why the extracted parameters do not retain the conventional meaning of local permittivity and permeability. These retrieved parameters, not considering the finite phase velocity within each unit cell, can strongly vary for different incidence angles and violate basic causality conditions.

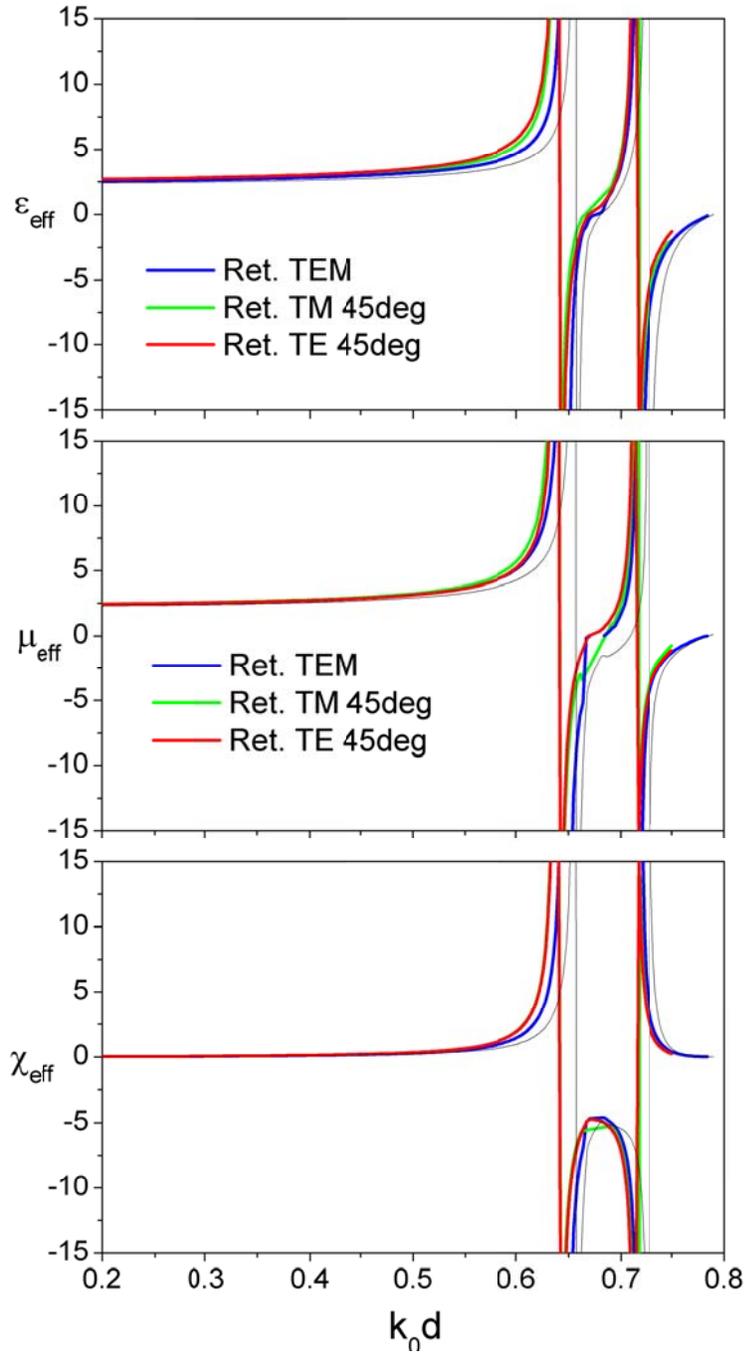

Figure 2. Effective parameters calculated using the analytical FPHT (thin-black line) and the proposed GRM for different incidence angles and polarizations.

We then apply the inversion procedure discussed in the previous section, solving the set of equations (2)-(4) and (10)-(11), with input $R$, $T$ and the additional information on the array period $d$, which we assume to know. Figure 2 shows the retrieved effective parameters $\varepsilon_{eff}$, $\mu_{eff}$, $\chi_{eff}^o$, with blue, green and red lines, referring to the results for normal incidence, 45-degree TE and TM incidence, respectively. The thin-black lines, reported for comparison, are the analytical results calculated using (2)-(4) and the Mie polarizability expressions for the isolated spheres, as reported in Ref. 27.

Quite excitingly, all the curves in Figs. 2a and 2b possess Lorentzian frequency dispersion, and they do not show negative slopes within the propagation region. The differences between retrieved results for different excitation angles become much smaller, and the residual discrepancy is simply associated with the large array density, which makes the coupling between neighboring spheres dependent on the field polarization, introducing some small form of anisotropy.

The retrieved curves of permittivity and permeability, consistent with the analytical ones, are remarkably different from the NRW retrieval. Each of them shows two distinct Lorentzian resonances, associated with the electric and magnetic resonance of the spheres, which affect both the electric and magnetic bulk response of the array, due to the properly captured magneto-electric coupling in the lattice. These effects are completely overlooked in the conventional retrieval results shown in Fig. 1.

It is possible to see a small frequency shift between retrieved curves and the analytical results. The obvious reason behind this shift lies in the dipolar approximations used in the direct FPHT model, in which we only take into account the dipolar coupling among neighboring inclusions. It is obvious that a full retrieval approach is able to capture the correct coupling among inclusions with much better accuracy, including higher-order multipolar interactions, and essentially provides a corrected polarizability response that includes near-field effects. These differences would be less important for arrays with smaller filling ratios and with weaker near-field coupling among neighboring inclusions, for which the retrieved parameters would converge more closely to the analytical curves. It is quite remarkable how the proposed GRM may essentially provide even more accurate results than our analytical derivation, going beyond the dipolar approximation used in Ref. 20.

Figure 2c shows the retrieved $\chi^o_{eff}$, which highlights the relevance of the magneto-electric coupling near the bandgap and in the negative-index region. This quantity is completely neglected in conventional retrieval procedures, but is evidently responsible for the deviations from a Lorentzian, causal dispersion of the *equivalent* permittivity and permeability in Fig. 1. Away from the inclusion resonance frequency where the array performs more as a mixture, these effects are negligible, i.e. $\chi^o_{eff} \sim 0$, and the results converge to a conventional retrieval.

As outlined above, the GRM also returns, as a byproduct of the inversion procedure, the electric and magnetic polarizabilities of the inclusions. Figure 3 shows the retrieved values, compared with the analytically calculated Mie polarizability coefficients for the spheres considered in this example. It is seen how all the retrieved curves are perfectly Lorentzian in shape, satisfy causality, passivity and the radiation condition[31]. Consistent with the previous results, the resonance frequency of our retrieved polarizabilities is slightly red shifted compared to the analytical Mie polarizability, due to the near-field coupling between neighboring spheres. It is interesting that even in the retrieved curves different field polarizations predict slightly shifted inclusion resonances, due to the modification of near-field coupling. These effects disappear for less close spheres.

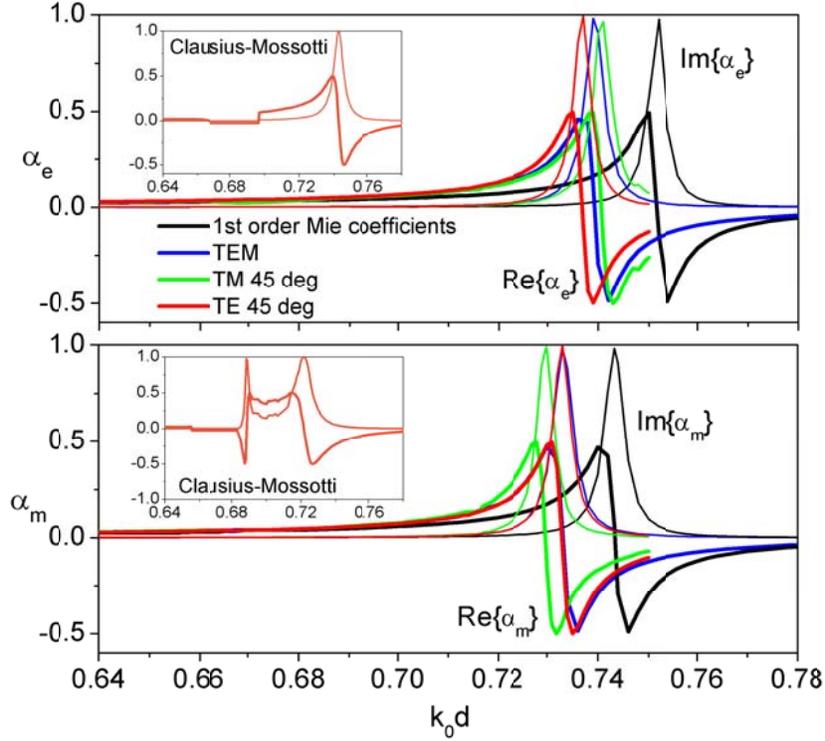

Figure 3 - Electric and magnetic polarizability obtained from analytical Mie theory, our generalized retrieval method and a conventional Clausius-Mossotti model.

It is striking to compare these retrieved polarizability curves not only with the expected analytical results for the sphere polarizability as in Fig. 3, but also with the values we would get using a more conventional Clausius-Mossotti model[32, 33] that neglects the magneto-electric coupling. We obtain these *equivalent* polarizabilities of the array by assuming again to know $d$ and inverting the equations

$$\varepsilon_{eq} = \varepsilon_0 \left( 1 + \frac{d^{-3}}{\alpha_e^{-1} - C_{int}} \right), \quad (14)$$

$$\mu_{eq} = \mu_0 \left( 1 + \frac{d^{-3}}{\alpha_m^{-1} - C_{int}} \right), \quad (15)$$

obtained from (2)-(3) after neglecting the magneto-electric interaction $C'_{em}$. These are shown in the insets of Fig. 3, highlighting the nonphysical, non-Lorentzian frequency dispersion that is implicitly assumed in the inclusions when we adopt a more conventional retrieval approach. As we get closer to the metamaterial resonance, and over a quite broad range of frequencies, it is obvious that a simplistic constitutive model that neglects the magneto-electric coupling stemming from the lattice granularity would completely fail to describe the inclusions as causal and passive particles and cannot be considered a physically meaningful description of the array. Our retrieval procedure, on the contrary, returns physically meaningful bulk material parameters and the polarizability of the inclusions composing the array under analysis, correctly capturing these hidden magneto-electric effects.

Finally, it is important to test how sensitive our method is to the knowledge of the array period. In practical configurations, in fact, the array may not be ideally periodic, and/or the information on the average distance among the inclusions may not be known in precise terms. For this reason, in Fig. 4, we show the retrieved parameters obtained after running our retrieval algorithm for different period values $d_r$ within a $\pm 5\%$ variation from the correct value. We stress that in calculating these results we are keeping the scattering parameters $R$ and $T$ fixed, based on the numerical results used in the previous calculations, and we assume the same total thickness of the sample to extract the secondary parameters. As evident in the figure, our results are very stable after changing this input parameter. The Lorentzian nature of the curves is nicely preserved, as it is fundamentally at the basis of the physical mechanisms considered in the FPHT. Slight differences are noticed, as expected, only near the bandgap, where $\chi^o_{eff}$ is more relevant. For larger values of $d_r$ the effective parameters get slightly smaller in this region and the Lorentzian resonance shifts to slightly larger frequencies.

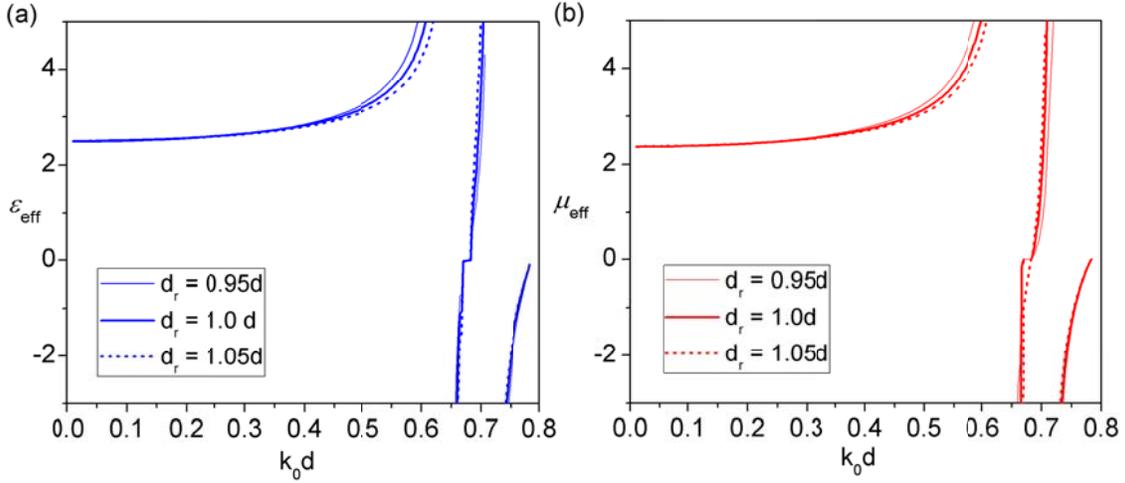

Figure 4 - Effective parameters obtained by the GRM with different values of lattice period based on the same set of scattering coefficients.

IV. **Conclusions**

In this work, we have proposed a generalized retrieval method for periodic metamaterials based on an extension of the NRW retrieval procedure combined with our recently reported FPHT. This approach takes into account the full dynamic coupling in dense periodic lattices to determine physically meaningful effective parameters, which have been shown to satisfy fundamental principles, such as Kramers-Kronig conditions, even in frequency ranges for which conventional retrieval methods would inherently fail. Compared to conventional retrieval, this method requires the knowledge, to a reasonable degree, of the average period or distance between neighboring inclusions, to extract the weak spatial dispersion effects arising near the metamaterial resonances. As a byproduct, we are also able to retrieve the electric and magnetic polarizabilities of the unit cells, which have been shown to be consistent with causality and passivity conditions, different from what we would get using more conventional approaches. We have applied this novel retrieval approach to an array of magnetodielectric spheres forming a negative-index metamaterial and confirmed the physical nature of the retrieved parameters. This method represents an important advancement in the characterization of metamaterials and can address many of the significant challenges experienced so far in providing physical meaning to the retrieved parameters of metamaterial samples. By using this technique to test metamaterial samples, it may be possible to restore the physical meaning of the retrieved effective permittivity and permeability of metamaterials, and describe these arrays as bulk, homogeneous materials independent of the form of excitation.

*Acknowledgements*

This work has been supported by the ONR MURI grant No. N00014-10-1-0942. We would like to thank Prof. Nader Engheta for useful suggestions and discussions.

**References**

1. D. R. Smith, J. B. Pendry and M. C. K. Wiltshire, Science **305**, 788792 (2004).
2. J. Huangfu, L. Ran, H. Chen, X. Zhang, K. Chen, T. M. Grzegorczyk, and J. A. Kong, Appl. Phys. Lett. **84**, 1537 (2004).
3. D. R. Smith, S. Schultz, P. Markoš and C. M. Soukoulis, Phys. Rev. B **65**, 195104 (2002).
4. X. Chen, T. M. Grzegorczyk, B.-I. Wu, J. P., Jr., and J. A. Kong, Phys. Rev. E **70**, 016608 (2004).
5. X. Chen, B.-I. Wu, J. A. Kong, and T. M. Grzegorczyk, Phys. Rev. E **71**, 046610 (2005).
6. J. Baker-Jarvis, E.J. Vanzura, and W.A. Kissick, - IEEE Trans. Microwave Theory Tech. **38**, 1096-1103 (1990).

7. C. R. Simovski, "On electromagnetic characterization and homogenization of nanostructured metamaterials," *J. Opt.* **13,** 013001 (2011).
8. J. Woodley and M. Mojahedi, JOSA B **27**, 1016-1021 (2010).
9. D. R. Smith, J. G. Gollub, J. J. Mock, W. J. Padilla, and D. Schurig, J. Appl. Phys. **100**, 024507 (2006).
10. X.-X. Liu, D. A. Powell, and A. Alù, Phys. Rev. B **84**, 235106 (2011).
11. P. A. Belov and C. R. Simovski, Phys. Rev. E **72**, 026615 (2005).
12. D. R. Smith and J. B. Pendry, JOSA B **23** 391-403 (2006).
13. M. G. Silveirinha, Phys. Rev. B **75**, 115104 (2007).
14. J. D. Baena, L. Jelinek, R. Marqués, and M. Silveirinha, Phys. Rev. A **78**, 013842 (2008).
15. C. Fietz and G. Shvets, Physica B **405** 2930–2934 (2010).
16. R. A. Shore and A. D. Yaghjian, Radio Sci. **42**, RS6S21 (2007).
17. A. Pors, I. Tsukerman, and S. I. Bozhevolnyi, Phys. Rev. E **84**, 016609 (2011).
18. O. Ouchetto, C.-W. Qiu, S. Zouhdi, L.-W. Li, and A. Razek, IEEE Trans. Microwave Theory Tech. **54**, 3893 – 3898 (2006).
19. S. Zhou, W. Li, and Q. Li, IEEE Trans. Microwave Theory Tech. **58**, 910-916 (2010).
20. A. Alù, Phys. Rev. B **84**, 075153 (2011).
21. A. Alù, Phys. Rev. B **83**, 081102(R) (2011).
22. H. Wallén, H. Kettunen, J. Qi, and A. Sihvola, General Assembly and Scientific Symposium, 2011 XXXth URSI.
23. In this paper, we use the words 'effective' and 'equivalent' referring to the context introduced in [20,21]. Equivalent parameters do not retain the physical meaning of local permittivity and permeability, but only describe the bulk scattering properties of the sample. The effective parameters, on the contrary, are physical quantities that have local properties in the long-wavelength limit.
24. M. G. Silveirinha, Phys. Rev. B **76**, 245117 (2007).
25. M. G. Silveirinha, C. A. Fernandes, IEEE Trans. Antennas Propag. **53**, 347-335 (2005).
26. A. H. Sihvola, A. J. Viitanen, I. V. Lindell and S. A. Tretyakov, Electromagnetic Waves in Chiral and Bi-Isotropic Media (Artech House Antenna Library, 1994).
27. X.-X. Liu and A. Alù, Metamaterials **5**, 56-63 (2011).
28. X.-X. Liu and A. Alù, J. Nanophoton. **5,** 053509 (2011).
29. C. Menzel, C. Rockstuhl, T. Paul, F. Lederer, and T. Pertsch, Phys. Rev. B **77**, 195328 (2008).

30. It should be noted that strictly speaking for oblique incidence, due to lack of symmetries, the propagation in the metamaterial cannot be necessarily modeled as a purely transverse electromagnetic (TEM) mode, as implicitly assumed in (1), (5)-(6), and a suitable tensorial notation should be used to properly capture the propagation mechanisms. In this subwavelength limit, however, it is possible to prove that the TEM assumption implicitly made in this paper holds to a very good degree, justifying the present discussion.
31. A. Alù, A. D. Yaghjian, R. A. Shore, and M. G. Silveirinha, Phys. Rev. B **84**, 054305 (2011).
32. J. E. Sipe and J. Van Kranendonk, Phys. Rev. A **9**, 1806–1822 (1974).
33. S. A. Tretyakov, *Analytical Modeling in Applied Electromagnetics* (Artech House, London, 2002).